\documentclass[12pt]{article}
\usepackage{mathrsfs}
\usepackage{amsmath}
\usepackage{mathrsfs}
\usepackage{amssymb}
\usepackage{color}
\usepackage{psfrag}
\usepackage{amsmath}
\usepackage{amssymb}
\usepackage{graphicx}%\documentclass[prd,aps,preprint]{revtex}

\newcommand{\bea}{\begin{eqnarray}}
\newcommand{\eea}{\end{eqnarray}}
\newcommand{\be}{\begin{equation}}
\newcommand{\ee}{\end{equation}}
\newcommand{\vs}[1]{\vspace{#1 mm}}

\newcommand{\dsl}{\pa \kern-0.5em /}

\newcommand{\pa}{\partial}

\newcommand{\nn}{\nonumber\\}

\begin{document}
\topmargin 0pt
\oddsidemargin 0mm

%\renewcommand{\thefootnote}{\fnsymbol{footnote}}
%\begin{titlepage}
\begin{flushright}
USTC-ICTS-04-13\\
\end{flushright}

\vs{0.5}
\begin{center}
{\Large \bf Dynamical brane creation and annihilation \\
\vspace{2mm}
 via a background flux} \vs{10}

{\large Yi-Fei Chen and J. X. Lu
 }

 \vspace{5mm}

{\em
  Interdisciplinary Center for Theoretical Study\\
 University of Science and Technology of China, Hefei, Anhui 230026,
P. R. China\\
%and\\
%$^2$ Interdisciplinary Center of Theoretical Studies\\
%Chinese Academy of Sciences, Beijing 100080, China\\
 %and\\

%$^3$ Michigan Center for Theoretical Physics\\
%Randall Laboratory, Department of Physics\\
%University of Michigan, Ann Arbor, MI 48109-1120, USA\\
}
\end{center}

\vs{5}
\centerline{{\bf{Abstract}}}
\vs{5}
\begin{small}
We study the dynamical Myers effect by allowing the fuzzy (or the
dynamical dielectric brane) coordinates to be time dependent. We
find three novel kinds of the dynamical spherical dielectric
branes depending on their respective excess energies. The first
represents a dynamical spherical brane carrying a negative excess
energy (having a lower bound) with its radius oscillating
periodically between two given non-zero values. The second is the
one with zero excess energy and whose time dependence can be
expressed in terms of a simple function. This particular dynamical
spherical configuration represents the dielectric brane creation
and/or annihilation like a photon in the presence of a background
creating an electron-position pair and then annihilating back to a
photon. The third is the one carrying positive excess energy and
the radius can also oscillate periodically between two non-zero
values but, unlike the first kind, it passes zero twice for each
cycle. Each of the above can also be interpreted as the time
evolution of a semi-spherical D-brane--anti semi-spherical D-brane
system.
\end{small}
\newpage

The discovery of static SUSY preserving BPS p-branes in string
theory helps establishing various duality relations among
different string theories and the eleven dimensional supergravity
and leads to the unification of these theories to a big not yet
established theory called M-theory. This latter theory is
non-perturbative in nature and as such our understanding of it is
so far still very limited, in spite of much progress made during
the last two decades,  mainly due to our lack of ability in dealing
with non-perturbative phenomenon in general. It is fair to say
that the current efforts in string/M theory community are largely
still in the stage of theoretical data collection for the
M-theory. Any thing non-trivial and going beyond the static BPS
configurations and related is in general hard and worth trying.
Among these, the studies of  D-brane--anti D-brane systems and the
related tachyon condensations made by Sen and others stand
out\cite{asone,astwo}. The recent astronomical observation of the
accelerating expansion of our universe which leads to a small
positive cosmological constant requires also the effort in
string/M theory community to study possible dynamical processes
which might provide an explanation of this.

   We here try to make a small step in this direction by studying
the dynamical creation and/or annihilation of a higher dimensional
spherical D-brane from lower dimensional seed BPS D-branes via a
R-R flux in a given background. This is motivated by the
following: Myers studied the creation of a static dielectric brane
from neutral objects like D0 branes in the presence of 4-form RR
flux, the analog of the static dielectric effect in ordinary
electromagnetism. Can such a dynamical brane be created and what
are its properties and the corresponding physical effect? We know
that a photon with sufficient energy can create an
electron-positron pair and then annihilates back to a photon but
this requires the presence of a background field to conserve the
energy-momentum. Can we realize such an analog in the present
case?  As we will see, we can indeed do so in addition to many
other novel physical effects. This work provides not only a
dynamical process of the dielectric brane creation and/or
annihilation but also a non-trivial concrete realization of the
dynamics of a Dp-brane --anti Dp-brane system. This work might also provide
means for viable cosmological model building.

   To make the basic ideas behind the current work clear, we in
this paper limit our study on generating a spherical D2 brane from
$N$ BPS seed D0-branes via a constant R-R four-form flux. In
addition to the R-R flux, we will consider Minkowski flat metric
which satisfies the bulk equations of motion to the leading order
once certain constraints (mentioned later on) are satisfied  for
large $N$. Extending the current work to other branes and other
backgrounds is straightforward.

   Myers made a remarkable observation a while ago\cite{myersone}
that lower dimensional D-branes can couple to the RR potentials of higher ranks, associated  with  higher
dimensional D-branes, if the non-Abelian effect of these lower
dimensional D-branes is taking into consideration. As such, he
found that a static charge neutral (but carrying a R-R
electric-like dipole moment) spherical D2-brane can be produced
from the D0-branes in the presence of a constant R-R 4-form flux
background\footnote{The magnetic analogue of this was later
discussed in \cite{dtvone}.}. We here try to extend Myers
and others' work and find that Myers static spherical D2-brane is
 a special case of a large class of  dynamical spherical
D2-branes and there are rich dynamical properties associated with
them. Actually there are three kinds of these dynamical branes
depending on the excess energies above the original D0-branes for
which we now discuss one by one in order.

     To make things concrete, let us consider first the matrix theory
description of N D0 branes in the presence of a constant RR 4-form
flux and a flat bulk spacetime with $G_{\mu\nu} = \eta_{\mu\nu},
B_{\mu\nu} = 0, \phi = {\rm constant}$, following from\cite{myersone,dtvone}. The
corresponding Lagrangian up to the order of ${\cal O} (\lambda^2)$
is \be \label{lag} L = - T_0 N  + \frac{T_0 \lambda^2}{2} {\rm Tr}(\dot\Phi^i)^2 +
\frac{T_0 \lambda^2}{4} {\rm Tr}\, [\Phi^i, \Phi^j]^2 + \frac{i
\lambda^2 T_0}{3} {\rm Tr}\Phi^i \Phi^j \Phi^k F_{tijk}.\ee Here we
choose already the static gauge $t = \sigma^0$, $\lambda = 2\pi
l_s^2$ with $l_s$ the string length scale, $T_p = 2\pi/[g_s (2\pi
l_s)^{p + 1}]$ the Dp-brane tension with $g_s = e^\phi$ the string
coupling. As in \cite{myersone}, we take \be F_{tijk} = \left\{
\begin{array} {ll} - 2 f \epsilon_{ijk} & \mbox{for $i, j, k \in \{1, 2, 3\}$}\\
0&\mbox{otherwise}\end {array} \right. ,\ee with $f$ carrying
dimensions of $\mbox{length}^{-1}$. The non-trivial part of the
equations of motion along $1, 2, 3$ directions from the above
Lagrangian is \be \ddot \Phi^i + [[\Phi^i, \Phi^j], \Phi^j] + 2 i
f  \Phi^j \Phi^k \epsilon_{ijk} = 0,\ee where we have made use of
(2) for $F_{tijk}$. As in \cite{myersone}, we take $\Phi^i$ for $i
= 1, 2, 3$ in the $N\times N$ irreducible representation of
$SU(2)$ but instead we are seeking a general dynamical charge
neutral (but carrying a time dependent RR dipole moment) spherical
D2 brane configuration. In other words, we take $\Phi^i = u(t)
J^i$ with $u(t)$ a function of time in general and the $SU(2)$
generators $J^i$ satisfying $[J^i, J^j] = i \epsilon_{ijk} J^k$.
Then the above equations are reduced to a single second order
non-linear differential equation \be\label{ueqn} \ddot u - 2 f u^2 + 2 u^3 =
0,\ee which can be derived either from the following action  
\be\label{rlag} L = - T_0 N + \frac{\lambda^2 T_0 N(N^2 - 1)}{8} \left({\dot u}^2 - u^4 + \frac{4}{3} f u^3\right),\ee
or the  Hamiltonian 
 \be \label{hamiltonian}H = T_0 N + \frac{\lambda^2 T_0 N(N^2 - 1)}{8} \left({\dot u}^2 + u^4 - \frac{4}{3} f u^3\right). \ee
 In having the above, we have set $\Phi^i = u (t) J^i$ for $i = 1, 2, 3$ and $\Phi^i $ to be diagonal constant matrices for  $i \neq 1, 2, 3$, for simplicity\footnote{In general, these $\Phi^i$ for $i \neq 1, 2, 3$ can have a motion of the center of mass.}, in the original action (\ref{lag}). We have also made use of
  \be
  {\rm Tr} (J^i)^2 = \frac{1}{12} N (N^2 - 1), 
 \ee
 for $i = 1, 2, 3$, respectively.  We can read from either (\ref{rlag}) or (\ref{hamiltonian}) the following effective potential for $u$ as 
 \be \label{potential} V = \frac{\lambda^2 T_0 N(N^2 - 1)}{8} \left(u^4 - \frac{4}{3} f u^3\right).\ee  Figure 1 shows the characteristic behavior of the above potential with two extremal points at $u = 0$ and $u = f$, respectively,  for which the former is in fact an (decreasing) inflection point while the latter gives the minimum of the potential. In addition to $u = 0$, the other zero point of the potential occurs at $u = 4 f/3$ as is evident from the expression of the potential.
\begin{figure}
\centering
\includegraphics[scale=0.6]{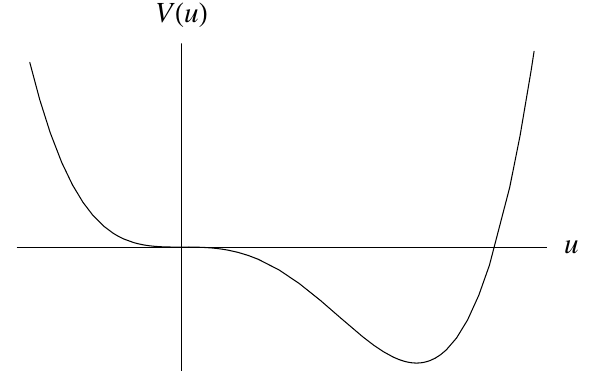}
\caption{The characteristic behaviour of potential $V (u)$ vs $u$}
\end{figure}
 
 Similar to the static case discussed in\cite{myersone}, we
expect that the equation (\ref{ueqn}) can also be derived from the
worldvolume action of a single spherical dynamical D2 brane with a
wordvolume $U(1)$ flux describing $N$ D0 branes $F_{\theta\phi} =
N \sin\theta /2$ moving in a flat Minkowski background with the
coordinates $x^i$ for $i = 1, 2, 3$ expressed in spherical polar
ones as \be ds^2 = - dt^2 + dr^2 + r^2 (d\theta^2 + \sin^2\theta
d\phi^2) + \sum_{i = 4}^9 (d x^i)^2.\ee In doing so, we also take
the static gauge $\sigma^0 = t, \sigma^1 = \theta, \sigma^2 =
\phi$ and take the radial coordinate $r$ to be time dependent and
$x^i$ ($i = 4, \cdots, 9$) to be constant. We further need to take
the motion to be non-relativistic, i.e., $\dot r \ll 1$ and $r^2
\ll \lambda N/2$ which can indeed be satisfied consistently as we
will see later on. Notice also now that $F_{tr\theta\phi} = - 2f
r^2 \sin\theta \rightarrow C_{t\theta\phi} = 2 f r^3 \sin\theta
/3$. With all these, the equation of motion for $r$ can indeed be
reduced to the one for $u(t)$ above if we identify $r (t) =
\lambda N |u(t)|/2$. So $r \ll \sqrt{\lambda N/2}$ implies that $u
(t) \ll \sqrt{2/(\lambda N)}$ which further implies $ f \ll
{\cal O} (1/\sqrt{\lambda N})$ once the solution for $u (t)$ is considered.

  Let us comment a few things before we proceed. In general, our
  chosen flat background with a constant RR 4-form flux does not
  satisfy the bulk equations of motion as mentioned in
  \cite{myersone}. The equivalence of the above two
  descriptions puts constraints on both $u(t)$ and the parameter
  $f$. In particular, for large $N$, both $u(t)$ and $f$ are tiny but the size of
  the spherical D2 brane can still be large since we need only $r \ll \sqrt{\lambda
  N/2}$ as pointed out in \cite{myersone}. The smallness of $u(t)$
  makes our ignoring higher order terms in the matrix action
  justified while that of $f$ implies that the flat background
  satisfies the bulk equations of motion to the leading order since
  now the metric correction due to the presence of the
  4-form $\delta h_{\mu\nu} \ll 1$  as discussed in \cite{dtvone}. We therefore
  should take large $N$ to validate our following discussion.

  Our
  focus now is equation (4). Integrating it once, we have
  \be \label{inte}  \dot u^2 = C + \frac{4}{3} f u^3 - u^4,\ee
  where the integration constant $C$ is actually the so-called reduced  excess
  energy $\Delta E_r \equiv 8 (H - T_0 N)/ (\lambda^2 T_0 N (N^2 - 1)) =  {\dot u}^2 + u^4 - 4 f u^3/3$ of the system as can be seen from the Hamiltonian (\ref{hamiltonian}) and is no less than $- f^4/3$ from the above equation given that $\dot u^2(t) \ge 0$ and the reduced potential $V_r \equiv 8 V /(\lambda^2 T_0 N (N^2 - 1)) = u^4 - 4 f u^3/3$ has a minimum value of $- f^4/3$ at $u = f$. Note that $\dot u = 0$ gives the extremes of the $u(t)$
  with respect to time $t$.  Except for $C = - f^4/3$ corresponding to Myers static
  solution with $\dot u =0, \ddot u = 0$ and $u = f$, and $C = 0$ corresponding to a degenerate
  root $u (t) = 0$ and the other $u (t) = 4 f/3$, we have in general four roots of the quartic equation obtained by setting $\dot u = 0$  in (\ref{inte}) and
  two of them denoted as $a$ and $b$ with $a > b$ are real and the other two are complex conjugate to
  each other denoted as $c$ and $\bar c$. These roots can be expressed in terms of $f$ and $C$ as,
  \bea \label{roots} 
  &&a = \frac{f}{3} \left[1 + y^{1/2}  + \sqrt{3 + 2 y^{-1/2} - y} \right],\quad  b = \frac{f}{3} \left[1 + y^{1/2}
  - \sqrt{3 + 2 y^{-1/2} - y}\right], \nn 
   && c = \frac{f}{3} \left[1 - y^{1/2} - i \sqrt{y + 2 y^{- 1/2} - 3} \right], \eea
   where we have $y = 1 + 3 x^{1/3} [(1 - \sqrt{1 - x})^{1/3} + (1 +
  \sqrt{1 - x})^{1/3}]/2$ for $- f^4/3 \le C < 0$ with $0 < x \equiv - 3 C/f^4 \le 1$ , and $y = 1 - 3 x^{1/3}[(\sqrt{1 + x} + 1)^{1/3} - (\sqrt{1 + x} -
  1)^{1/3}]/2$  for $C > 0$ with $x \equiv 3 C/f^4 > 0$. From (\ref{inte}), we then have $\dot u^2 (t) = - (u - a) (u - b) (u - c)(u - \bar c) \ge 0$, giving 
  $ b \le u (t) \le a$. We then have the general solution in terms of Jacobian elliptic function $\mbox{cn} (t)$ as
   \be \label{solution} 
  u(t) =\frac{a B + A b - \left(a B - A b\right) \mbox{cn} \left(\frac{t - t_0}{g}\right)}{
  \left(A + B\right) + \left(A - B\right)\mbox{cn}\left(\frac{t - t_0}{g}\right)},\ee
  where $A^2 = (a - b_1)^2 + a_1^2$ and $B^2 = (b - b_1)^2 + a_1^2$ with $a_1^2 = - (c - \bar
  c)^2/4,\, b_1 = (c + \bar c)/2,\, g = 1/\sqrt{AB}$. 
     In addition to Myers' static solution, we have from the above the other three kinds of
  time dependent solutions, depending on the value of the integration constant $C$.  We now come to discuss each of them in order.
  
  Case (1):  $ - f^4/3 \le C < 0$.  Except for $C = - f^4/3$ giving Myers' static solution $u  = f$ the D2-anti D2 ground state,  the general solution in this case 
  cannot be expressed in terms of
  elementary functions and one has to use Jacobian elliptic function $\mbox{cn} [(t - t_0)/g]$ with $1/g = 2f (2 y^2 + 1/y - 3 y)^{1/4}/3$. Since the $\mbox{cn} x$
  is a periodic function like $\cos x$, taking values also between $-1$ and $1$ with its standard period (the corresponding
  $k^2  = 1/2 + 3(1 - y)/[4 (2 y^2 + 1/y - 3y)^{1/2}]$), therefore
  $u(t)$ is also a periodic function oscillating between the roots
  $b$ and $a$ as given in (\ref{roots}) with respect to its expected equilibrium point at $u = f$ which can be determined, for example, by the requirement of maximum speed from (\ref{inte}) or the minimum of the potential (\ref{potential}). Note that since both $a$ and
  $b$ are positive for the present case due to $1 < y \le 4$, $u(t)$ is also positive which
  implies that the radius of the spherical D2 brane oscillates
  periodically between $\lambda N b/2$ and $\lambda N a/2$ with its equilibrium radius $ r = \lambda N f/2$.  According to \cite{gmtone}, a spherical dielectric
  D2-brane can be viewed as a semi-spherical D2-brane --anti semi-spherical D2-brane system and as such one would naively expect the 
  appearance of a tachyon mode. However, the equilibrium position $r = \lambda N f/2$ of this brane-anti brane occurs at  the potential minimum of the N D0 system at $u = f$ and this indicates that the D2 brane should be stable if the back-reaction as well as the
  radiation are ignored. In other words, we should not expect a tachyon mode to appear in this case(or the would-be tachyon mode becomes a stable one).   We will discuss this more later on.

  Case (2): for $C = 0$, i.e., the total excess energy is zero, in addition to the trivial solution
  $u (t) = 0$, we have the following non-trivial one
   \be u (t) = \frac{12 f }{9 + 4
  f^2 (t - t_0)^2},\ee with $t_0$ the integration constant. We can
  interpret this solution in two ways depending on whether we take our initial
  time $t = -\infty$ or $t = t_0$. If we take $- \infty < t <
  \infty$, then at initial and final stages, $u(t) = 0$, $\dot
  u(t) =  0,\,$ and $ \ddot u(t) = 0$,
  therefore we have only N D0 branes at the two ends while in-between we have the
  creation of the dynamical spherical D2-brane. This is the analog of
  a photon creating an electron-positron pair and then annihilating
  back to a photon but the current process takes an infinity amount of
  time and the number of D0 branes is large. Interpreting the spherical D2 as
  a semi-spherical D2--anti semi-spherical D2 pair and
   with the 4-form flux switching on,
  we create this pair whose size increases with time up to $t =
  t_0$ where the size reaches its maximum and then decreases to
  zero at $t\to\infty$. If we choose our initial time at $t = t_0$, then the pair
  starts from the largest size and ends with a zero one, becoming N D0 branes. Again, we don't expect the appearance of unstable tachyon mode since neither the extremal $u = 0$ nor the extremal $u = f$ point of the potential $V (u)$ given in (\ref{potential}) gives rise to any tachyon mode. 
  We will elaborate more on this later on.

 Case (3): for $ C > 0$, i.e., with a positive excess energy,
  the solution now for $u(t)$ has exactly the same form as
  in case (1) as given in (\ref{solution}) and once again we need
  to use the Jacobian elliptic function. But in this case,
  the variable $y$ is different as given below (\ref{roots}).  The other
  difference is that now the parameter $b < 0$ due to $0 < y < 1$. Since $u(t)$ oscillates
  periodically between $b$ and $a$ ($a > f$), so we expect that $u(t)$ goes
  through zero twice for each cycle. In other words,
  assuming initial zero size for the spherical D2 branes (or just N D0 branes)
  the radius of the D2 brane grows with time, for example, passing through the equilibrium radius at $\lambda N f/2$ to reach the
  maximal size $\lambda N a /2$, then shrink to zero size again after passing through the equilibrium radius one more time, then
  grows toward the size $\lambda N |b|/2$ and finally shrinks back
  to zero size to finish one cycle. One can also interpret this
  process as the time evolution of a semi-spherical D2 brane--anti
  semi-spherical D2 brane with positive excess energy, but once again one does not expect the appearance of unstable tachyon mode for the same reason as given before. 
  We would like to point out here about the connection of
  the present solution or the one in case (1) to the one in case (2).
  Actually, if we take $C \to 0$ from the solution either in case
  (1) or in case (3), it reduces to either the trivial $u (t) = 0$ or the non-trivial
  one given by (7) in case (2). For the former case, the Jacobian elliptic
  function $\mbox{cn} [(t - t_0)/g] = 1 - (t - t_0)^2/(2 g^2)$ while for
  the latter $\mbox{cn} [(t - t_0)/g] = - 1 + (t - t_0)^2/(2 g^2)$ where
  for each case $g \to \infty$ as $C \to 0$ with all the other parameters
  in the solution also take proper limits.

  In each of the above three cases, the excess energy is conserved
  because we take the system under consideration  basically as an isolated one
  (we ignore radiation and back reaction). Because of this, the
  system in case (1) has less total energy than that of N D0
  branes  and therefore the spherical D2 brane cannot shrink its
  size to zero (or the semi-spherical D2--anti semi-spherical D2 pair
  cannot annihilate each other back to N D0 branes) to set down
  to N D0 branes. Case (2) has zero
  excess energy which is just the right amount of excess energy for
  the final state of N BPS D0 branes. Therefore, we must expect
  that the final state consists of N D0 branes as the solution indicates.  In case (3), we
  have a positive excess energy above N BPS D0 banes which has no way to go
  given our assumption of no interaction with surroundings. Now
  the spherical D2 brane
  has enough energy to shrink its size to
  zero (or the D-brane--anti D-brane pair has enough energy to annihilate each
  other) but it cannot stay there because of the
  positive excess energy.  Given our assumption, the dynamical D2 brane in each case as well as the N D0 branes in case (2) has no way to give away its or their excess energy above the ground state to settle down to the true Myers' static spherical D2 or the D2-anti D2 ground state.

     In practice, the system under consideration has couplings
     with the bulk and with other branes even though these could  be
     weak for large $N$ and $\lambda N  f^2 \ll 1$. We therefore have radiation
     and/or back-reaction for such system. Note that the $u = 0$, corresponding to the true ground state of N D0 branes in the absence of the RR 4-form flux, is now an inflection point while Myers' blown-up static spherical D2 brane or the D2-anti D2 is the true ground state which has $u = f$ and occurs  at the potential minimum. Therefore all the dynamical motion considered in this paper can be viewed one way or the other as the one oscillating around the equilibrium position $u = f$ or the D2-anti D2 ground state.  This implies that all the dynamical D2 branes as well as the D0 branes will eventually settle down to the static spherical D2 or the D2- anti D2 ground state after the underlying system radiates away their excess energy above the ground state via interactions with other modes. This may happen via a chain of processes such as from case (3) to case (2), then to case (1) and finally to the ground state or directly from case (3) to case (1), then finally to the ground state and so on,  depending on where we start with and how the energy is radiated.  Here we must assume that the interactions involved must be weak and the ground state itself is not altered in the radiation process.         
     
     Given what has been said, for large $N$ and small flux $\lambda N  f^2 \ll 1$,  the interactions of the underlying system with all other possible modes should be weak enough and the system itself is therefore almost isolated. We then expect that each kind of dynamical solutions discussed in this paper should live with a rather long lifetime. This is further supported by the absence of unstable tachyon mode.  At first look, this is rather strange since each of the dynamical D2 is actually a D2-anti D2 system and in general we should expect the appearance of a tachyon mode. However, at either extremal $u = 0$ or extremal $u = f$ point of the potential (\ref{potential}), we don't have a tachyon mode. The former corresponds to an inflection point of the potential, indicating the possible instability of the adjoint coordinate modes associated with N D0 branes. For this,  we expect that the N D0 branes, in particular the solution $ u = 0$ for these D0 branes, are unstable and the adjoint coordinates associated with them are condensing.  The latter corresponds to the point of potential minimum, giving rise to Myers' static spherical D2 or the D2-anti D2 ground state. As discussed in \cite{senthree,astwo,senfour, yione}, whether a brane-anti brane system is stable (or a tachyon mode appears) depends on certain parameter(s)  associated with the system.  Before switching on the 4-form RR flux, the N D0 branes are BPS stable one.  After switching on, the flux only lowers the potential minimum from zero to a negative one and the system new ground state becomes more stable than the original N D0 one even though it is a D2-anti D2. This very fact indicates also the absence of unstable tachyon mode following the line in \cite{senthree,astwo,senfour, yione}. Since the ground state is stable,  we don't expect that the dynamical D2, viewed as excited one in each case, has an issue with the tachyonic instability.

     If we replace the 4-form RR flux by the flux $H_{ijk}$ as
     pointed out in \cite{myersone}, we will end up with the same
     action, therefore the same conclusion can be reached.

     Note added: Just about submitting the initial version of the present work to
     arXiv, we received a preprint \cite{rstone} in which a different time dependent D2 brane solution is discussed.

\vs{5}

\noindent {\bf Acknowledgements}

\vs{2}

We acknowledge support by grants from the Chinese Academy of
Sciences and the grants from the NSF of China with Grant
No:11235010,10245001, 90303002.

%%%%%%%%%%%%%%%%%
%%%%%%%%%%%%%%%%%
%%%%%%%%%%%%%%%%%


\begin{thebibliography}{99}

\bibitem{asone} A. Sen, ``Non-BPS states and branes in string theory'',
hep-th/9904207.

\bibitem{astwo} A. Sen, ``Tachyon condensation on the brane-antibrane
system'',
JHEP 08 (1998) 012, [hep-th/9805170].

\bibitem{myersone} R. C. Myers, ``Dielectric branes", JHEP 9912
(1999) 022, [hep-th/9910053].

\bibitem{dtvone} S. R. Das, S. P. Trivedi and S. Vaidya,
``Magnetic moments of branes and giant gravitons", JHEP 0010
(2000) 037, [hep-th/0008203].

\bibitem{gmtone} M. T. Grisaru, R. C. Myers and O. Tafjord, ``SUSY
and Goliath", JHEP 0008 (2000) 040, [hep-th/0008015].

\bibitem{senthree} A. Sen, ``Stable non-BPS bound states of BPS
D-branes", JHEP 9808 (1998) 010, [hep-th/9805019].

\bibitem{senfour} A. Sen, ``SO(32) spinors of Type I and other solitons on
brane-antibrane pair", JHEP 9809 (1998) 023, [hep-th/9808141].

\bibitem{yione} P. Yi, ``Membranes from five-branes and fundamental strings
from Dp branes", Nucl.Phys. B550 (1999) 214, [hep-th/9901159].

\bibitem{rstone} S. Ramgoolam, B. Spence and S. Thomas,
``Resolving brane collapse with $1/N$ corrections in non-abelian
DBI", [hep-th/0405256].

\end{thebibliography}
\end{document}